\documentclass[aps,pra,twocolumn,showpacs,superscriptaddress,groupedaddress]{revtex4-1}  % for review and submission
\usepackage{graphicx}  % needed for figures
\usepackage{dcolumn}   % needed for some tables
\usepackage{bm}        % for math
\usepackage{amssymb}   % for math
\usepackage{overpic}
\usepackage[small]{caption}

\usepackage{wasysym}
\usepackage{color}
\usepackage{float}
\usepackage[labelsep=space,labelfont=bf,justification=raggedright]{caption}
\newcommand{\ddd}{\, \mathrm{d}} 
\newcommand{\ci}{\mathrm{i}} 
\newcommand{\cl}{\mathrm{c}} 
\newcommand{\vekn}[1]{\mathrm{\textbf{#1}}}

\begin{document}
% the following line is for submission, including submission to the arXiv!!
%\hspace{5.2in} \mbox{Fermilab-Pub-04/xxx-E}

\title{Impact of slow-light enhancement on optical propagation in active semiconductor photonic crystal waveguides}
\author{Yaohui Chen, Jakob Rosenkrantz de Lasson, Niels Gregersen and Jesper M{\o}rk}
\email{jesm@fotonik.dtu.dk}
\affiliation{DTU Fotonik, Department of Photonics Engineering, Technical University of Denmark, \O rsteds Plads, Building 343, DK-2800 Kongens Lyngby, Denmark}

\begin{abstract}

We derive and validate a set of coupled Bloch wave equations for analyzing the reflection and transmission properties of active semiconductor photonic crystal waveguides. In such devices, slow-light propagation can be used to enhance the material gain per unit length, enabling, for example, the realization of short optical amplifiers compatible with photonic integration. The coupled wave analysis is compared to numerical approaches based on the Fourier modal method and a frequency domain finite element technique. The presence of material gain leads to the build-up of a backscattered field, which is interpreted as distributed feedback effects or reflection at passive-active interfaces, depending on the approach taken. For very large material gain values, the band structure of the waveguide is perturbed, and deviations from the simple coupled Bloch wave model are found.
 
\end{abstract}

\maketitle

\section{I. Introduction}

Photonic crystal (PhC) structures have been proposed as a waveguide infrastructure for high-density photonic integrated circuits \cite{McNab:2003OE, Krauss:2003PSS, Notomi:2011IET}. Optical amplification is one of the fundamental functionalities, required for  compensating attenuation and coupling losses and thus scaling-up the number of integrated devices \cite{Kish:2011JSTQE}. A major advantage of  using photonic crystal waveguides, as opposed to standard ridge-type waveguides, when realizing active structures with embedded semiconductor gain layers, is the use of slow-light effects to enhance light-matter interactions. Thus, the spatial gain coefficient may be increased by exploiting slow-light propagation, enabling the realization of short devices suitable for photonic integration \cite{Dowling:1994JAP,Baba:2008NP, Ek:2014NC}. Similarly, nonlinear effects induced by carrier depletion may also be enhanced, leading to devices with ultra-low saturation power, which may be of interest for optical signal processing applications \cite{Chen:2013OE}.

\begin{figure}[htb]
\centering
\includegraphics[width=0.48\textwidth]{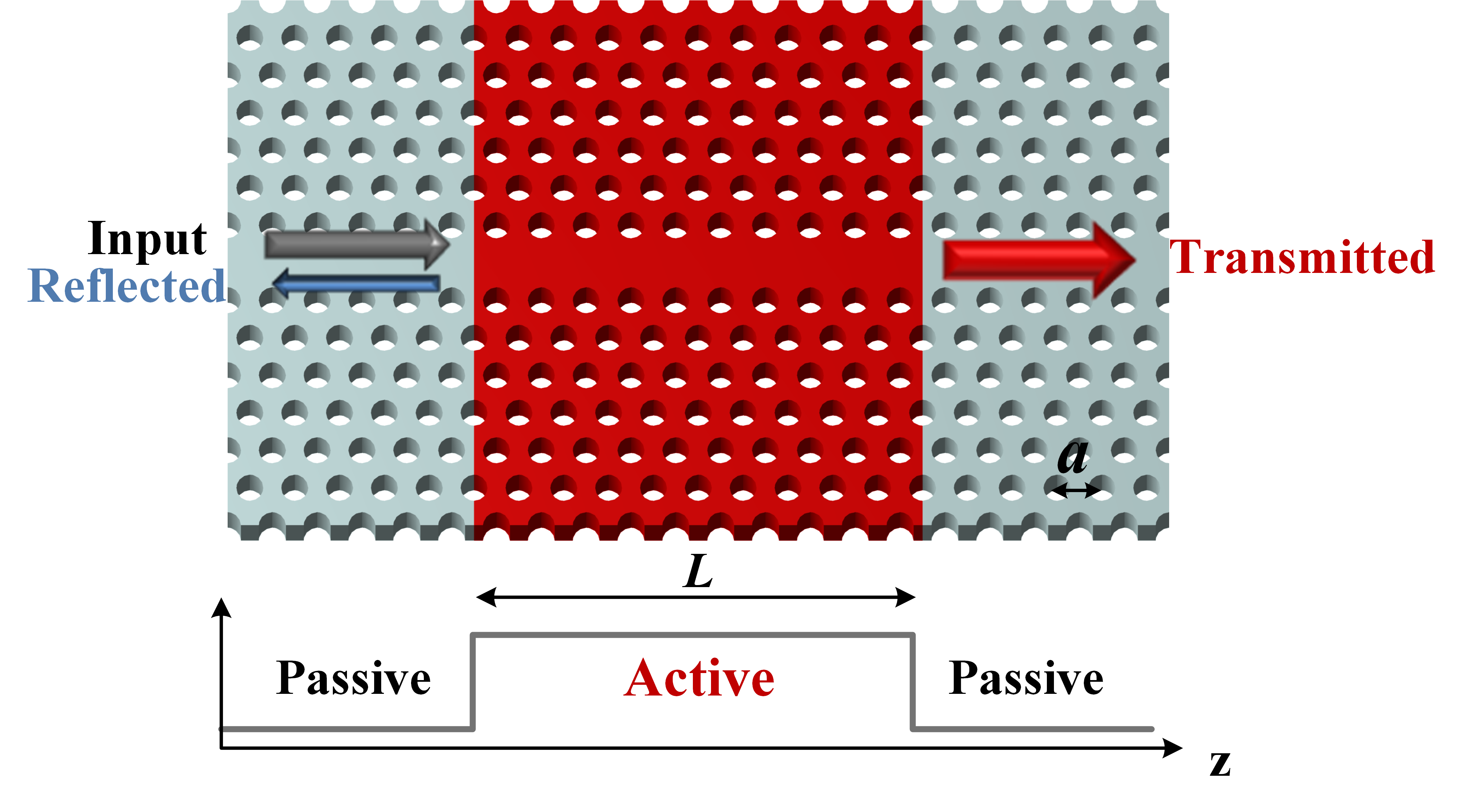}
\caption{Illustration of defect photonic crystal waveguide. The red part of the structure is active, i.e., this part of the membrane structure  contains embedded layers of quantum wells or quantum dots that may provide gain upon optical or electrical pumping. The lattice constant is $a$ and the length of the active region is $L$.}
 \label{fig:Limitation_1}
\end{figure}

In this work, we theoretically investigate the frequency-domain optical propagation properties of an active PhC waveguide of finite length embedded in an ideal passive periodic PhC waveguide platform as shown in Fig.~\ref{fig:Limitation_1}. Such structures with site-controlled active gain sections may be fabricated using different techniques \cite{Matsuo:2010NP,Nozaki:2012NP,Kuramochi:2014NP, Calic:2011PRL}.  

As shown in Fig.~\ref{fig:Limitation_1}, we  approximate the device as a photonic crystal heterostucture \cite{Istrate:2006RMP} with a slowly-varying envelope of the imaginary part of the refractive index (bottom part), and a fast periodic variation that naturally arises from the penetration of the air holes through the gain layer (top part). Such an envelope approximation in photonic crystal devices is analogous to the treatment of semiconductor opto-electronic devices.  Based on a perturbative approach \cite{Sipe:2000PRE}, effective one-dimensional (1D) coupled-wave analysis has been widely used to investigate the impact of slow light effects on optical properties of passive PhC waveguides, e.g. for efficient taper design \cite{Johnson:2002PRE}, Kerr nonlinearities \cite{Sipe:2004PRE,Panoiu:2010JSTQE}, and disorder-induced scattering \cite{Patterson:2009PRL, Mazoyer:2009PRL}. 

For perfectly periodic PhC waveguides, neglecting Kerr nonlinearity and disorder, a rigorous set of equations for the amplitudes of forward and backward propagating unperturbed Bloch waves may be derived \cite{Chen:2012IPR}. In such a formulation, the presence of active material in a finite section of the PhC waveguide leads to multiple scattering, which represents material-gain-induced coupling between the forward and backward Bloch waves of passive structures.  Such \emph{distributed feedback} (DFB) effects have so far not been considered in slow-light enhanced semiconductor optical amplifiers \cite{Mizuta:2006JJAP,Mork:2010OL, Grgic:2012PRL,Ek:2014NC}.  We show here that it is important to consider the impact of such feedback effects when calculating the gain of an active PhC waveguide. 

Alternatively, we may treat the structure as multiple waveguide sections, namely an active PhC waveguide section interfaced with two semi-infinite passive PhC waveguides on both sides, with distinct sets of Bloch modes in the passive and active PhC waveguide sections.  In this picture, the active-passive interfaces induce reflections and Fabry-Perot effects are important in determining the strength of the signal transmitted from the input to the output waveguide as well as the back-reflected signal.  We solve the optical multiple scattering problem in aggregates of generalized thin films using the Fourier modal method (FMM) (or Rigorously coupled-wave analysis (RCWA))  \cite{Noponen:1994JOSAA,Lecamp:2007OE,Li:1996JOSAA}, which has been widely implemented for various periodic photonic structures. Using these methods, Bloch waves are computed and sorted in both passive and active PhC sections~\cite{NoteBlochSorting}. 
 
In the following, we numerically validate that both approaches are equivalent in analyzing the proposed active PhC devices. A similar discussion for active Bragg grating structures can be found in \cite{Yariv:1975APL}. We also analyze the situation with very large material gain values, in which case the band structure of the photonic crystal structure is modified. 
  
We shall here limit our attention to the simplest type of photonic crystal waveguide, with one row of missing holes (W1-waveguide). In that case, slow-light propagation is realized close to the Brillouin zone edge and the bandwidth over which significant slow-down effects are realized is limited. It was shown, though, that by dispersion engineering the waveguide, a slow-light region of significant bandwidth can be obtained for a frequency range displaced from the band-edge \cite{Frandsen:2006OE}. A systematic approach for optimizing the group index bandwidth product was recently demonstrated \cite{Wang:2012JOSAA}. For such dispersion engineered structures, the large group index region may be centered around an inflection point of the dispersion curves and it was shown that strong losses (or gain) can induce drastic changes of the dispersion curves \cite{White:2012PRA}.
  
The paper is organized as follows: In Section II we derive the coupled Bloch wave equations. Section III presents the numerical results. We first consider a long waveguide with relatively weak gain, and compare the coupled wave analysis to numerical results obtained using the FMM. Secondly we consider a shorter waveguide, enabling the consideration of larger absolute gain coefficients as well as the use of a finite element numerical technique. Finally, the main conclusions are summarized in Section IV.

\section{II. Effective one-dimensional Coupled Wave Analysis: Perturbation and Distributed Feedback}

\begin{figure}[htb]
 \centering
 \includegraphics[width=0.4\textwidth]{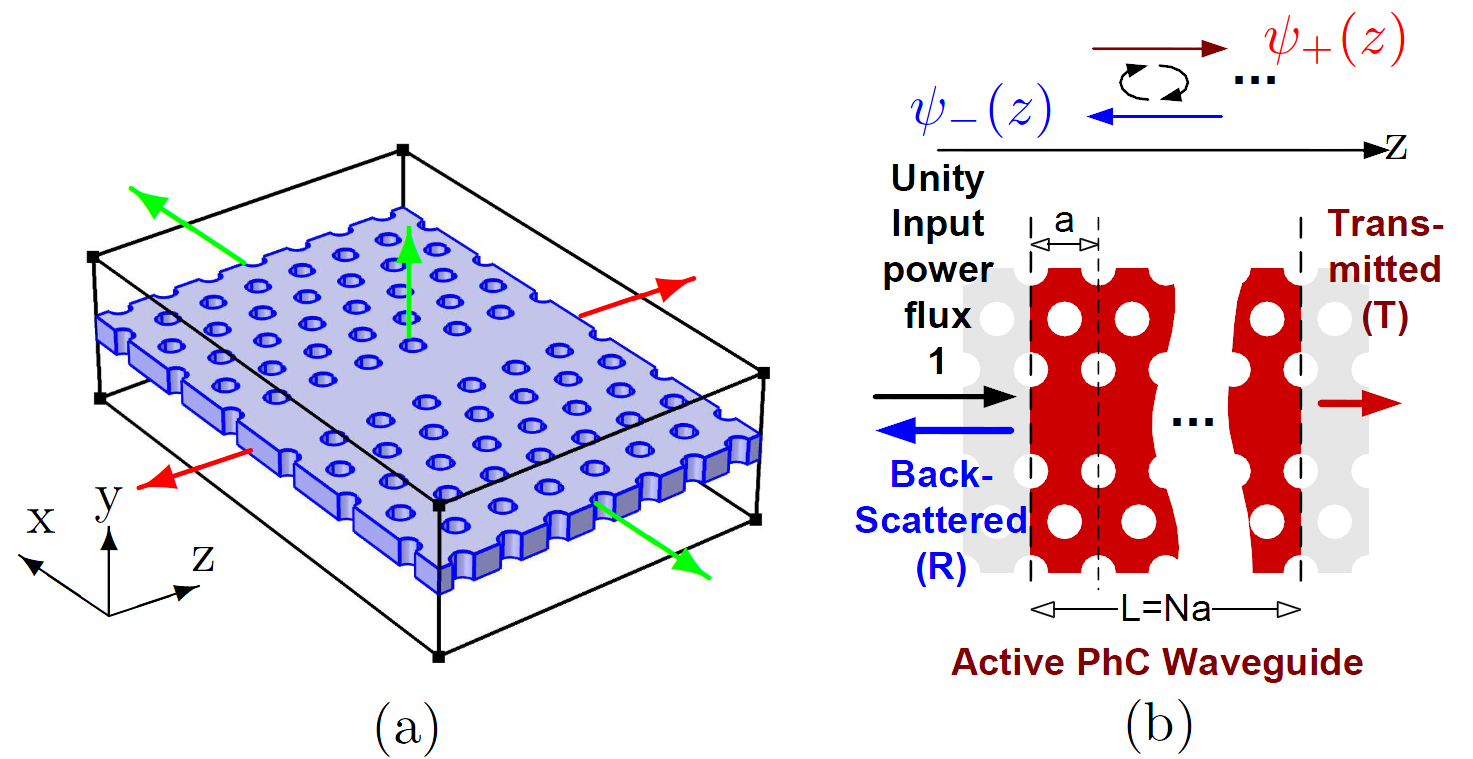}
\label{PhCstructure}
 \caption{Illustration of coupled wave analysis of PhC waveguide. (a) Schematic of surface used in the derivation of coupled mode equations, with arrows indicating outward unit normal vectors to the boundary surface. (b) Schematic diagram of coupling between the amplitudes of forward and backward propagating Bloch waves induced by the active material, which is represented by an imaginary perturbation of the refractive index.}
  \label{fig:Limitation_2}
 \end{figure}

We consider a line-defect photonic crystal waveguide, where part of the waveguide is active, as illustrated in Fig.~\ref{fig:Limitation_1}. The active material may be implemented as layers of quantum wells or quantum dots embedded in the middle of the photonic crystal membrane. In order to limit the energy consumption, it is preferable only to have active material in the core of the waveguide, as demonstrated in \cite{Matsuo:2010NP}. Here, however, we shall consider the simpler structure, where the membrane is uniformly active in the transverse directions ($xy$). This is appropriate for the case of optically pumped structures, as considered, for example, in \cite{Ek:2014NC}. We represent the material gain as a weak perturbation to the passive periodic PhC waveguide in a finite-length section and use the Bloch modes of the passive structure to expand the field of the structure including gain.  In this formulation, multiple scattering in the active PhC structure is represented as material-gain-induced coupling between the forward and backward propagating Bloch waves of the passive structure, as illustrated in the top part of Fig.~\ref{fig:Limitation_2}(b). This is similar to the way disorder is analyzed in \cite{Patterson:2009PRL}.
 
In the continuous wave (CW)  limit, the Maxwell equations can be rewritten based on the conjugated form of the Lorentz reciprocity theorem \cite{Michaelis:2003PRE,Panoiu:2010JSTQE}:
\begin{eqnarray}\label{eq:continuity_1}
\nabla \cdot ( \mathbf{E} \times \mathbf{H}_{0,\pm}^*+\mathbf{E}_{0,\pm}^*\times \mathbf{H})=\ci\omega \mathbf{P_{pert}}\cdot \mathbf{E}_{0,\pm}^*.
\end{eqnarray}
Here, $\mathbf{E}_{0,\pm}$ and $\mathbf{H}_{0,\pm}$ are the guided Bloch wave solutions of Maxwell equations for the electrical and magnetic fields  obtained in the absence of electronic polarization perturbations $\mathbf{P_{pert}}$, with $\pm$ denoting forward/backward propagating fields. The fields $\mathbf{E}$ and $\mathbf{H}$ are the self-consistent solutions in the presence of electronic polarization perturbations, which arise due to the presence of active material, e.g.~leading to stimulated emission and thus gain of the propagating fields. We may represent the unperturbed Bloch waves as follows:
\begin{eqnarray}\label{eq:Bloch_1}
\mathbf{E}_{0,\pm}=\frac{1}{2}\mathbf{e}_{\pm}(x,y,z)\exp(\pm \ci k_z z)\exp(-\ci\omega t),
\\ \label{eq:Bloch_1H}
\mathbf{H}_{0,\pm}=\frac{1}{2}\mathbf{h}_{\pm}(x,y,z)\exp(\pm \ci k_z z)\exp(-\ci\omega t),
\end{eqnarray}
with $\mathbf{e}_{\pm}(x,y,z)$ and $\mathbf{h}_{\pm}(x,y,z)$ being the normalized complex amplitude Bloch functions, $\omega$ the frequency and $k_z$ the Bloch wave number along the length of the waveguide. 

The unit rms power flux $P_z$ and unit rms electric and magnetic stored energy $W$ in a supercell are given as
\begin{eqnarray}
P_z&=&\frac{1}{2}\int_S  \left[\Re\{\mathbf{e} \times \mathbf{h}^*\}\cdot \mathbf{\hat{z}} \right] \ddd S, \\
W&=&\frac{1}{4}\int_V\left[\varepsilon_0n_b^2(\vekn{r})|\mathbf{e}|^2+\mu_0 |\mathbf{h}|^2\right]\ddd V=\frac{an_{gz}}{\cl}P_z.
\end{eqnarray}
Here, $n_{gz} \equiv \cl/v_{gz}$ is the group index, with $\cl$ and $v_{gz} = \partial \omega / \partial k_z$ being the speed of light and the group velocity, $\varepsilon_0$ and $\mu_0$ are the vacuum permittivity and permeability, $n_{b}(\vekn{r})$ is the background refractive index, $a$ is the lattice constant of the photonic crystal, $S$ indicates the transverse plane at position $z$, and $V$ is the volume of a PhC supercell. 
 
In the limit of a weak electronic perturbation, we express the self-consistent solutions as superpositions of the guided Bloch waves for the passive structure, with slowly varying amplitudes  $\psi_{\pm}(z)$: 
 \begin{eqnarray}\label{eq:Bloch_2}
\mathbf{E}&\simeq&\psi_+(z) \mathbf{E}_{0,+}+\psi_- (z)\mathbf{E}_{0,-} + \mathrm{c.c.}, \\ \label{eq:Bloch_2H}
\mathbf{H}&\simeq& \psi_+(z) \mathbf{H}_{0,+}+\psi_-(z)\mathbf{H}_{0,-} +  \mathrm{c.c.}.
\end{eqnarray}

By using Eqs.~(\ref{eq:Bloch_1}), (\ref{eq:Bloch_1H}), (\ref{eq:Bloch_2}) and (\ref{eq:Bloch_2H}), Eq.~(\ref{eq:continuity_1}) can be formulated as a set of continuity equations for the forward and backward propagating amplitudes:
\begin{equation}\label{eq:continuity_2}
\pm \frac{1}{2}\nabla \cdot \left[ \psi_{\pm}(z) \Re\{ \mathbf{e} \times \mathbf{h}^*\} \right]=\ci \omega \mathbf{P_{pert}}\cdot \mathbf{E}_{0,\pm}^*.
\end{equation}
Eq.~(\ref{eq:continuity_2}) can be derived using the divergence theorem for a closed surface $S$ enclosing the simulation domain:
\begin{equation}\label{eq:divergence_1}
\pm \frac{1}{2}\oiint_S \left[\psi_{\pm}(z)\Re\{  \mathbf{e}\times \mathbf{h}^*\}\cdot \mathbf{\hat{n}}\right] \ddd S= \ci\omega \int_V (\mathbf{P_{pert}}\cdot \mathbf{E}_{0,\pm}^*) \ddd V.
\end{equation}
Here, $\mathbf{\hat{n}}$ is the outward unit normal vector of the closed surface depicted in Fig.~\ref{fig:Limitation_2}(a). This closed surface can be chosen conveniently to comply with the experimental geometry considered (near field measurement \cite{Gersen:2005PRL} or transmission measurement \cite{Tanaka:2004EL}), as well as the simulation domain and boundary conditions (BCs) (e.g.~periodic BCs \cite{Johnson:2001OE} or perfectly-matched layer (PML) BCs  \cite{Lecamp:2007OE}). The left-hand side (LHS) of Eq.~(\ref{eq:continuity_2}) describes the possible channels for the power flux in/out of the surface and the right-hand side (RHS) describes the source term. Eq.~(\ref{eq:divergence_1}) is in the form of a set of implicit integral equations.  Assuming that the considered modes are transversely confined, the surface integral (LHS of Eq.~(\ref{eq:divergence_1})) reduces to two integrals between the planes $z=0$ and $z=L$, described as: 
 \begin{eqnarray}\label{eq:simple_1}
&\hspace{-3cm}\pm \frac{1}{2}\oiint_S \left[\psi_{\pm}(z)\Re\{  \mathbf{e} \times \mathbf{h}^*\}\cdot \mathbf{\hat{n}}\right] \ddd S 
\nonumber\\
&\simeq \pm\int_0^L \partial_z \psi_{\pm}(z) \ddd z \times \frac{1}{2}\int_{S} \left[\Re\{  \mathbf{e} \times \mathbf{h}^*\}\cdot \mathbf{\hat{z}}\right] \ddd S.
 \end{eqnarray}

For simplicity, we describe the carrier-induced complex susceptibility perturbation as a product of a complex constant $ \chi_{pert}$ and an active material distribution function $F(\vekn{r})$:
\begin{equation}\label{eq:simple_2}
\mathbf{P_{pert}}=\frac{1}{2}\varepsilon_0\chi_{pert}F(\vekn{r}) \mathbf{E},
\end{equation}
where $F(\vekn{r}) = 1$ ($= 0$) in the active (passive) region.

We now obtain the coupled-mode equations from Eq.~(\ref{eq:divergence_1}) with substitution from Eqs.~(\ref{eq:simple_1})-(\ref{eq:simple_2}):
\begin{eqnarray}
\partial_z \psi_+(z)=\frac{\ci\omega }{\cl} n_{gz} \chi_{pert}\left[ \delta(z) \psi_++ \kappa^*(z)e^{-\ci 2 k_z z}\psi_-\right]\label{CMeq+},\\
\partial_z \psi_-(z)=-\frac{\ci \omega }{\cl}n_{gz}  \chi_{pert}\left[ \delta(z) \psi_-+  \kappa(z)e^{\ci 2 k_z z}\psi_+\right]\label{CMeq-},
\end{eqnarray}
with 
\begin{eqnarray}
\delta(z)&\equiv& \frac{a \int_S  \left[\varepsilon_0 F(\vekn{r}) |\mathbf{e}|^2\right] \ddd S}{ \int_V \left[ \epsilon_0n_b^2(\vekn{r})|\mathbf{e}|^2+\mu_0 |\mathbf{h}|^2\right] \ddd V},
\\
\kappa(z) &\equiv &\frac{a\int_S   \left[\varepsilon_0 F(\vekn{r}) \mathbf{e}\cdot \mathbf{e}\right] \ddd S}{ \int_V \left[ \epsilon_0n_b^2(\vekn{r})|\mathbf{e}|^2+\mu_0 |\mathbf{h}|^2\right] \ddd V}.
\end{eqnarray}
Here, $\delta(z)$ and $\kappa(z)$ denote the complex-valued propagation and backscattering coefficient induced by the perturbation due to the active material. The above 1D coupled wave equations are mathematically equivalent to a $2\times 2$ scattering matrix problem in a stack of thin films (at each $z$ coordinate) and readily solved numerically. 

We note the similarity of the above coupled mode equations with the coupled mode equations originally derived by Kogelnik and Shank for distributed feedback lasers \cite{Kogelnik:1972JAP}. However, while those equations describe the slow variation of the complex amplitudes of plane waves (see also 
\cite{Olivier:2003OE} for application to PhC waveguides), Eqs.~(\ref{CMeq+})-(\ref{CMeq-}) describe the slowly varying amplitudes of the {\em Bloch} waves of the corresponding passive PhC. This has the consequence that in the absence of a polarization perturbation, i.e., $\chi_{pert}=0$, the forward and backward Bloch waves do not couple but remain uni-directional propagating waves, as required. The presence of a perturbation, however, leads to coupling of the Bloch waves, since the material perturbation gives rise to distributed feedback. This is analogous to the scattering between Bloch waves induced by structural disorder \cite{Patterson:2009PRL}, although in that case the backscattering sites are randomly distributed.

In general, the harmonic term $\exp(\pm \ci 2 k_z z)$  leads to relatively low backscattering efficiency due to phase mismatch between forward and backward propagating Bloch waves. Propagation losses decrease the effective propagation length and further damp the backscattering. In passive PhC waveguides, one can integrate similar coupled-mode equations approximately by neglecting the distributed feedback terms as long as the backscattered power is weak \cite{Johnson:2002PRE}. On the other hand, in the case of positive gain, the effective length, over which wave interaction takes place, increases, and a significant backscattered wave may be built up, as we shall show here. 

Eq.~(12)-(13) must be supplemented by boundary conditions at the passive-active interfaces. The boundary conditions \cite{Tromborg:1994JQE} are:
\begin{eqnarray}
\psi_{+}(0)&=&r_1\psi_{-}(0)+\psi_0,\\
\psi_{-}(L)\exp(-i k_z L)&=&r_2\psi_{+}(L)\exp(i k_z L),
\end{eqnarray}
where $r_1$ and $r_2$ are the amplitude reflectivities of the left and right passive-active interfaces and $\psi_0=1$ indicates unity incident field from the left interface. 

Finally, we note that the coupled wave approach presented here permits the inclusion of disorder-induced losses \cite{Patterson:2009PRL, Mazoyer:2009PRL} and carrier dynamics \cite{Chen:2013OE}, which is beyond the scope of this paper. 

\section{III. Simulation Results}
For the numerical investigations, we consider a two-dimensional (2D) triangular lattice of air holes (hole radius $r = 0.25a$) embedded in a semiconductor membrane with a dielectric constant of $n_b^2=12.1$. The waveguide is a W1 line-defect with a single row of air holes omitted. The presence of material gain or absorption is modeled via an imaginary refractive index perturbation $n_i$, corresponding to a material gain $g_0=-2n_i\omega/\cl$. The corresponding  susceptibility perturbation is $\chi_{pert}=-n_i^2+\ci 2n_bn_i$. In practice, the value of the gain coefficient is controlled via the charge carrier density. We notice that the ratio between the real and imaginary part of the susceptibility change is close to zero, corresponding to a near-zero value for the linewidth enhancement factor. In the results to be presented, we shall quantify the gain via the imaginary part of the refractive index; as examples, $n_i=-10^{-3}$ corresponds to a material gain of $g_0 = 81.1$ cm$^{-1}$ at $\lambda_0=1550$ nm and $g_0 = 96.7$ cm$^{-1}$ at $\lambda_0=1300$ nm. 

In the following we separately consider two waveguide lengths in order to illustrate and discuss two different physical effects. Section A considers the case of relatively long waveguides, but weak material gain, where a strong backscattered signal may build up. This situation will be shown to be accurately modeled by the derived coupled Bloch mode equations. Section B considers short waveguides, but with strong material gain, in which case the perturbative treatment breaks down and the imaginary contribution to the refractive index changes the band structure.

\subsection{A. PhC waveguides with a long active section}

\begin{figure}[ht]
 \begin{overpic}[scale=0.7]{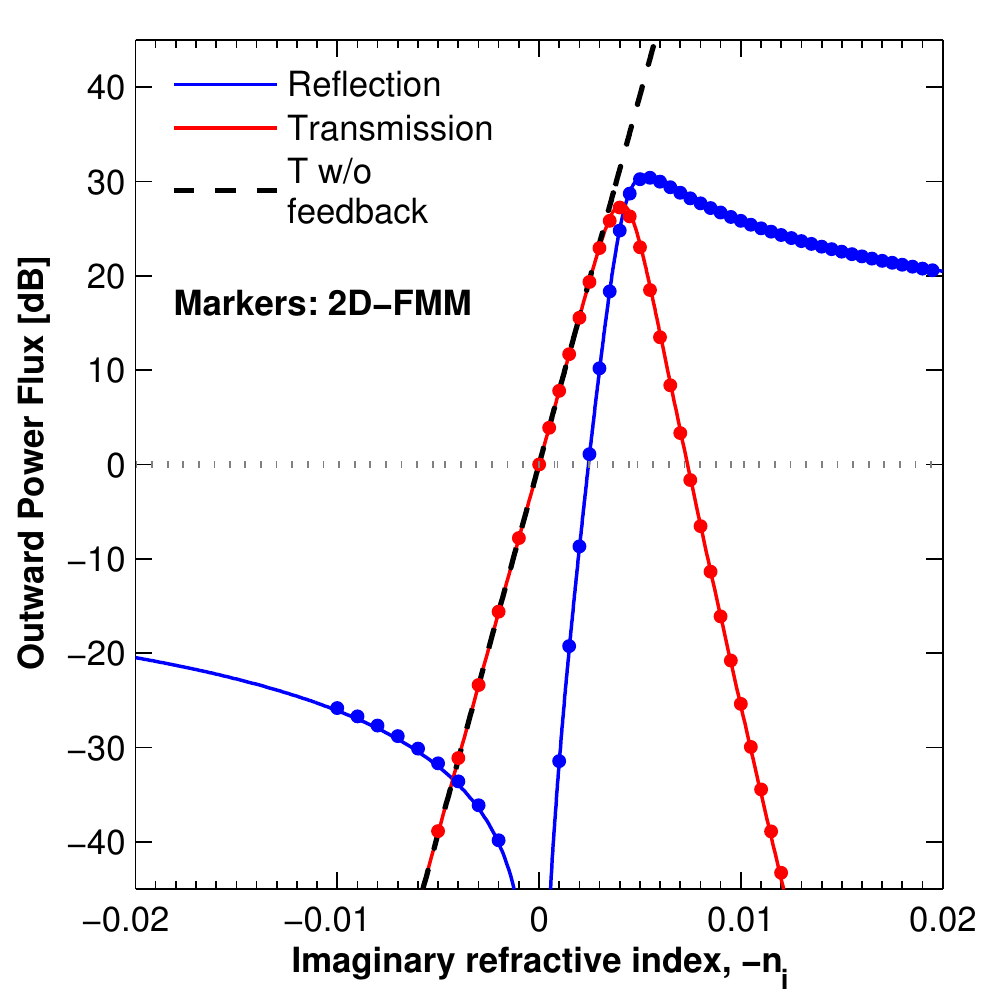}
 \end{overpic}
\caption{Power reflection and transmission for the PhC structure in Fig.~\ref{fig:Limitation_1} as functions of the imaginary part of the refractive index for a slow light mode at $\omega a/ (2\pi \cl)=0.2064$ with a group index $n_{gz} \simeq 25$ (see Figs.~\ref{fig:Limitation_4}(a) and (b)). The length of the active  waveguide is $L=100a$. Solid lines are results from the 1D coupled wave analysis, with $|r_1|^2=|r_2|^2=0$. Markers are 2D FMM simulation results.}
   \label{fig:Limitation_3}
\end{figure}
The length of the active PhC waveguide considered in this section is $L=100a$.
Numerically, such long waveguides are conveniently treated using the FMM that exploits the periodicity for obtaining the Bloch modes in both passive and active PhC sections. Specifically, it suffices to analyze a single unit cell (of length $a$), the eigensolutions of which are the Bloch modes~\cite{Lecamp:2007OE,NoteBlochSorting}, and the propagation through the full length ($L = 100a$) is afterwards handled analytically via Bloch's theorem. Active-passive reflection and transmission is handled using a scattering matrix approach~\cite{Lavrinenko2014}.

\begin{figure*}[htb]
\centering
\includegraphics[width=0.8\textwidth]{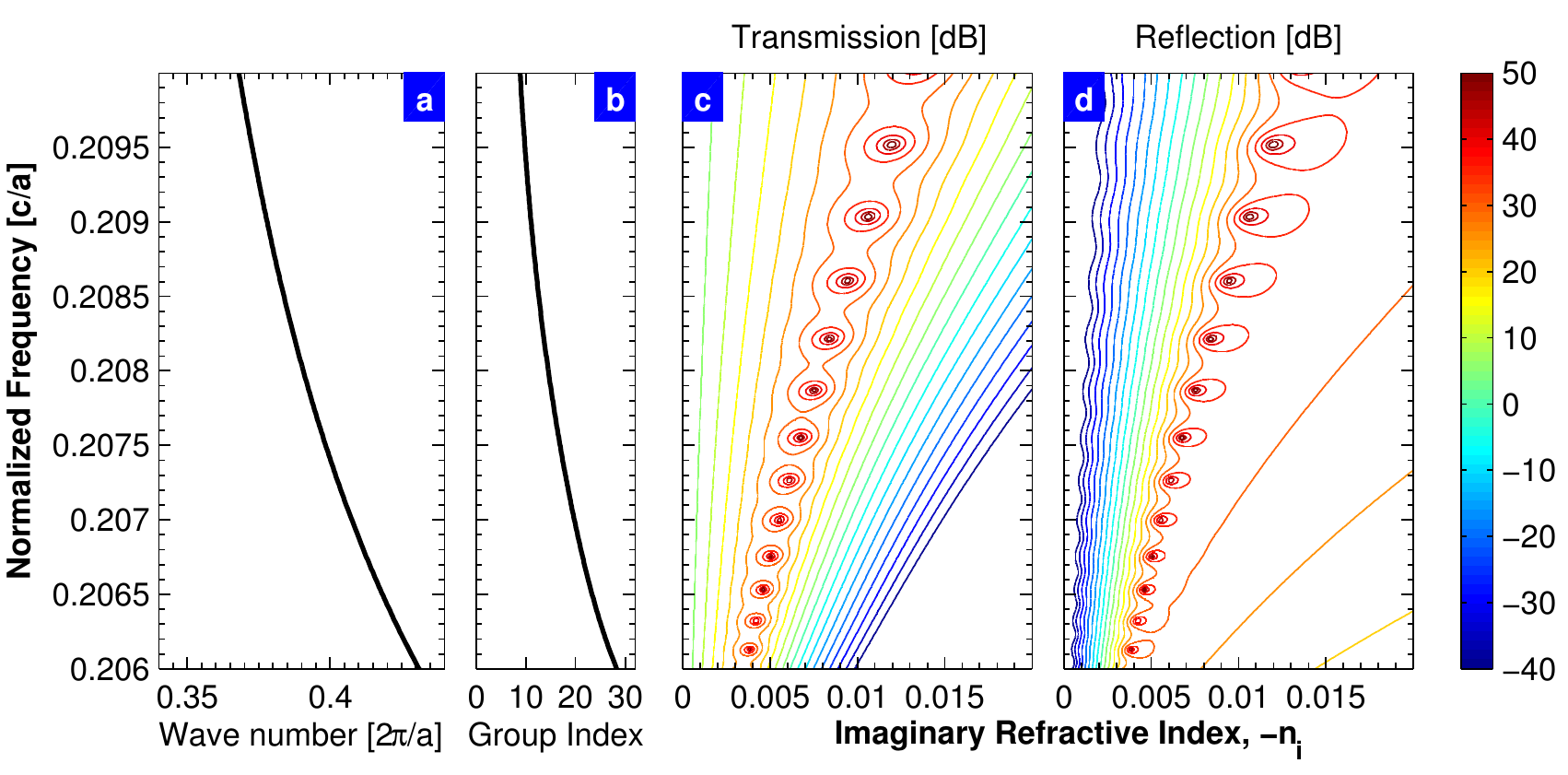}
\caption{Propagation characteristics of a line defect photonic crystal waveguide with a gain section of finite length, $L=100a$.  (a) Band diagram and (b) group index of passive PhC waveguide Bloch mode. (c) Power transmission and (d) reflection as function of the imaginary part of the refractive index and frequency. The results in (c) and (d) are obtained by numerically solving the 1D coupled Bloch wave equations.}
  \label{fig:Limitation_4}
\end{figure*}

Figure~\ref{fig:Limitation_3} shows the transmitted and reflected power flux of an input CW light exciting a slow light mode ($\omega a/ (2\pi \cl)=0.2064, n_g \simeq 25$, see Figs.~\ref{fig:Limitation_4}(a) and (b)) as functions of the imaginary part of the refractive index. 
 
The markers show the results based on 2D FMM simulations, while the solid lines are the results based on the solution of the 1D coupled Bloch  wave equations. As is apparent, the two approaches agree well. In both cases, a non-zero value of the imaginary part of the refractive index leads to the build-up of a reflected signal. The dashed line shows the ideal slow-light enhanced device gain/absorption \cite{Ek:2014NC}, $G= \exp(2n_{gz}n_bg_0\int_0^L\delta(z) \ddd z)=\exp(\Gamma \frac{n_{gz}}{n_b} g_0L)$, $\Gamma$ being the confinement factor, and the slope of which is proportional to the group index. It is seen that the simple slow-light enhanced gain expression in this case well accounts for the numerical results for absolute gain values up to about 25 dB. Beyond that the transmission starts to decrease, due to the build-up of a strong backreflected signal. 

In the absorption regime ($-n_{i}<0$), the increase  of  the material absorption leads to an increase of backscattered power flux, but the weak backscattered signal damped by material loss has negligible effect on the transmitted power.  On the other hand, in the regime of positive gain ($-n_{i}>0$), the forward as well as backscattered signals are amplified and the effective length over which the waves interact is increased. As the material gain increases, the transmitted power reaches a maximum value and then steeply decreases. 
  
We now briefly discuss the interpretation of the results in the context of the two different numerical techniques used, which effectively correspond to the use of different basis sets for representing the full solution. In the coupled Bloch mode approach, the total field is expanded on the Bloch modes of the passive waveguide (see Eqs.~(\ref{eq:Bloch_1}) and (\ref{eq:Bloch_2})). The inclusion of material gain or absorption corresponds to a periodic perturbation of the imaginary part of the refractive index and induces coupling between the passive Bloch modes (DFB) that are no longer solutions to the full, perturbed problem. In the FMM approach, the local solutions are instead expanded on the active Bloch modes, i.e., the solutions when including the imaginary index perturbation. For an infinitely long active waveguide, these solutions are uni-directional Bloch modes with exponentially growing powers, which renders this concept unphysical. In the case of a finite-length active waveguide, the appearance of a backreflected signal occurs due to the mismatch of the Bloch modes of the passive and active parts of the waveguide. This mode mismatch leads to  finite interface reflection coefficient and the appearance of Fabry-Perot (FP) resonances that depend on the active region length. The equivalence of the DFB and FP pictures was previously discussed in the context of simple 1D optical grating structures \cite{Yariv:1975APL}.  

Figures~\ref{fig:Limitation_4}(c) and (d) show the transmission and reflection as a function of frequency for varying imaginary part of the refractive index.  For a given frequency, as the imaginary part of the refractive index (material gain) increases, both transmission and reflection increase, reach a maximum and then start to decrease. For certain combinations of frequency and material gain both the transmitted and reflected beam diverge. These points correspond to the fulfillment of the lasing oscillation condition for a laser with length given by the active region. The corresponding frequencies correspond to discrete values of the wavenumber, with the real part of the wavenumber fulfilling the condition
\begin{equation}
 \left(\frac{\pi}{a}-\mathrm{Re}(k_z)\right)2  L=2\pi m,\,\,\,\,\,\, m = \mathrm{integer}.
\end{equation}
Taking into account saturation of the gain medium due to stimulated emission, the divergences correspond to the onset of self-sustained lasing, with the imaginary part of the refractive index being clamped to its threshold value. Again, we notice the similarity of these results with the classical analyses of  distributed feedback lasers \cite{Kogelnik:1972JAP} and active Bragg gratings \cite{Yariv:1975APL}.   
  
For frequencies closer to the bandgap, the resonances appear at smaller gain values, consistent with the enhancement of the gain per unit length due to slow light propagation \cite{Ek:2014NC}. Notice that the gain per unit time inside the material is not changed, with the consequence, for lasers, that it is only the contribution of mirror losses to the laser threshold that is decreased when exploiting a slow-light mode.

\subsection{B. PhC waveguides with a short active section }
    \begin{figure*}[!htb]
 \includegraphics[width=0.8\textwidth]{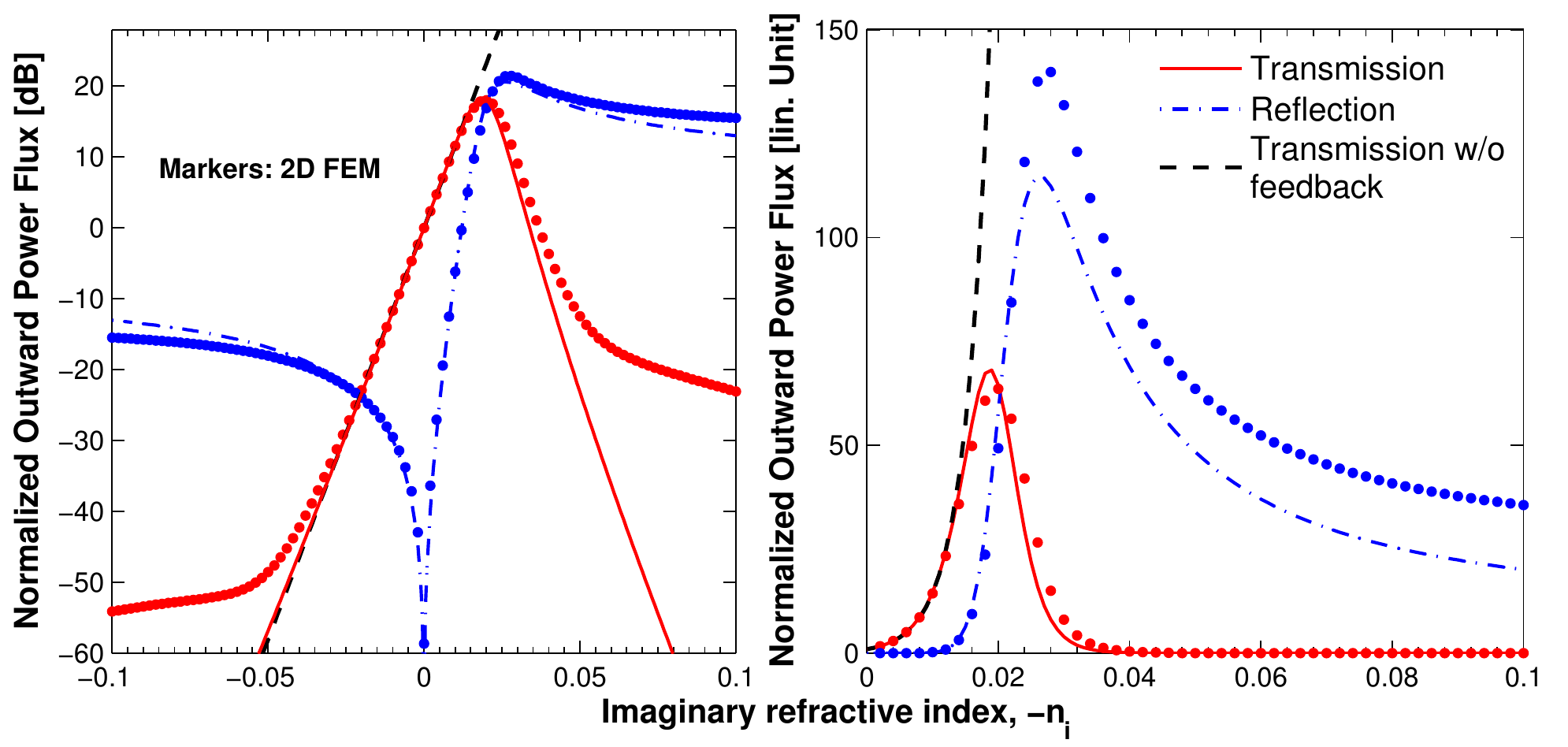}
\centering
\caption{Power transmission and reflection in (a) dB scale and (b) linear scale as functions of the imaginary part of the refractive index, for a short active  waveguide of length $L=20a$. The frequency is $\omega a/(2\pi c)=0.2075$, with a corresponding group index of $n_{gz}\simeq 18.7$.  Solid and dash-dotted lines are 1D coupled wave analysis results with zero passive-active interface reflections ($r_1 = r_2 = 0$). Markers are 2D FEM simulation results.}
    \label{fig:Limitation_5}
 \end{figure*}
In this section we consider a short active PhC waveguide section, of length $L=20a$, but allow for higher values of the material gain coefficient in order to explore the regime investigated in \cite{Grgic:2012PRL}.
As the size of 2D simulation domain decreases, it becomes feasible to solve the  periodic optical waveguiding problem \cite{Bao:1995SIAM} in the frequency-domain using the finite element method (FEM) with numerically exact Bloch mode excitation and/or absorptive boundary conditions \cite{Fietz:2013JOSAB}. 

Figure~\ref{fig:Limitation_5} shows the transmitted and reflected power flux of an input CW light exciting a slow light mode ($\omega a/ (2\pi \cl)=0.2075, n_{gz} \simeq 18.7$) as a function of the imaginary part of the refractive index. The solid and dash-dotted lines are obtained from the 1D coupled wave equations, while the markers are corresponding numerical results based on 2D FEM. The results are shown in both logarithmic and linear scales in order to appreciate the differences between the two approaches. We see that the coupled wave analysis reproduces most features seen in the full numerical FEM simulations for values of material gain/absorption coefficient up to 4000 cm$^{-1}$ ($|n_i|$ up to 0.05) at a wavelength of 1550 nm, which is large compared to gain values typically obtained in semiconductors. For smaller gain values, the shorter device shows characteristics similar to the longer device, with the absolute gain being smaller for the same gain coefficient. We attribute the difference between the coupled wave model and the full numerical result to a change of the effective gain coefficient appearing as the pre-factor of $\Psi_+$ and $\Psi_-$ in the RHS of Eqs.~(\ref{CMeq+})-(\ref{CMeq-}). Thus, for a  large absolute value of the imaginary part of the refractive index (large gain or absorption), the band structure is modified such that the group index is reduced and thereby also the slow-light enhanced gain and absorption. This is in agreement with the result obtained in \cite{Grgic:2012PRL}, where the band structure of an infinitely periodic structure was analyzed. In that case, there is no effect of backscattering, since a uni-directional Bloch wave of the active structure is considered, but consistent with the results presented here the slow-light enhancement was observed to decrease with the absolute value of the material gain coefficient.
      
We interpret the gain-induced change of effective group index (slow-light enhancement factor) as follows. The origin of the photonic crystal bandgap is the destructive interference of multiply scattered waves. However, in the presence of gain (or absorption), the destructive interference is incomplete. For a defect photonic crystal waveguide, this means that the band-edge is smeared and the group index no longer diverges when approaching the band-edge. In consequence, there is a gain-induced reduction of the group index, and thereby a gain-induced reduction of the slow-light enhancement of light-matter interaction. 

In the example above we considered a relatively modest group index of $\simeq 18.7$, but when approaching the band-edge further, the (passive) group index rapidly increases, and the saturation effect discussed above becomes more apparent, i.e., sets in at lower values of the absolute gain coefficient.

\section{IV. Conclusions}
We have derived a set of coupled Bloch wave equations that are useful for simulating and understanding the properties of active photonic crystal waveguides. The coupled Bloch wave model explains how slow-light effects can be used to enhance the absolute gain of a photonic crystal waveguide of given length, an effect recently observed experimentally \cite{Ek:2014NC}. The model also shows, however, that the presence of material gain (an imaginary refractive index perturbation) induces distributed feedback effects, additional to the slow light enhancement of modal gain, which leads to the build-up of a backscattered field that limits the attainable gain. The coupled Bloch wave model is compared to numerical simulations based on the Fourier modal method as well as a frequency domain finite element solver. In general, very good agreement between the different numerical approaches is found, although, depending on the approach used, the presence of a backscattered field calls for different, but consistent, interpretations. For very large values of the material gain (or absorption) coefficient, the band structure of the active photonic crystal itself is modified, leading to further changes of the slow-light enhancement, which is an effect not captured by the 1D coupled Bloch wave analysis. 

Finally, we notice that the transparent and computationally efficient formulation of the coupled Bloch wave equations presented here allows for the inclusion of additional relevant effects. In \cite{Patterson:2009PRL}, random disorder effects in passive structures were analyzed using a similar coupled Bloch wave analysis. With the recent observation of lasing in Anderson-localized modes due to disorder in active photonic crystal waveguides \cite{Liu:2014NN}, it appears natural to combine the two approaches. 

\section{Acknowledgements}
Yaohui Chen expresses his gratitude to Prof. Christian Hafner for helpful discussions on computational electromagnetics. Funding from Villum Fonden via the VKR Centre NATEC is gratefully acknowledged.

%\bibliographystyle{osajnl}  
%\bibliographystyle{abbrv} %The style doesnÕt matter
%\bibliography{Ref_PRL}

\end{document}